\documentclass[%
 reprint,
 amsmath,amssymb,
 aps,
]{revtex4-1}

\usepackage{graphicx}
\usepackage{dcolumn}
\usepackage{bm}
\usepackage{color}

\begin{document}

\preprint{APS/123-QED}

\title{Calculation of anisotropic transport coefficients for an ultrarelativistic Boltzmann gas in a magnetic field within a kinetic approach}

\author{Zhengyu Chen$^{1}$, Carsten Greiner$^{2}$, Anping Huang$^{1}$,}
\author{Zhe Xu$^{1}$}%
 \email{xuzhe@mail.tsinghua.edu.cn}
\affiliation{$^1$ Department of Physics, Tsinghua University and Collaborative Innovation Center of Quantum Matter, Beijing 100084, China}
\affiliation{$^2$ Institut f$\ddot{u}$r Theoretische Physik, Johann Wolfgang Goethe-Universit$\ddot{a}$t Frankfurt, Max-von-Laue-Strasse 1, 60438 Frankfurt am Main, Germany}

\date{\today}

\begin{abstract}
According to the Kubo formulas we employ the (3+1)-d parton cascade, Boltzmann approach of multiparton scatterings 
(BAMPS), to calculate  the anisotropic transport coefficients (shear viscosity and electric conductivity) for an ultrarelativistic
Boltzmann gas in the presence of a magnetic field. The results are compared with those recently obtained by using
the Grad's approximation. We find good agreements between both results, which confirms the general use of the derived
Kubo formulas for calculating the anisotropic transport coefficients of quark-gluon plasma in a magnetic field.
\end{abstract}

\pacs{Valid PACS appear here}
\maketitle


\section{Introduction}
\label{sec:introduction}
The experiments of ultrarelativistic heavy-ion collisions at the Relativistic Heavy Ion Collider (RHIC) and the Large Hadron
Collider (LHC) are believed to reach high enough energies to create the quark-gluon plasma (QGP) \cite{Gale:2013da},
which is composed of deconfined quarks and gluons at or close to thermal equilibrium. The QGP behaves like a nearly
perfect fluid with a small value of the shear viscosity to the entropy density ratio 
\cite{Adams:2005dq,Adcox:2004mh,Aamodt:2010pa,Gyulassy:2004zy,Arsene:2004fa,Back:2004je}.  
Due to the existence of strong magnetic field in the early stage of relativistic heavy-ion collisions 
\cite{Skokov:2009qp,Deng:2012pc,Tuchin:2013apa,Bzdak:2011yy}, the QGP may behave different from the one when
ignoring the magnetic field. Actually, the magnetic field breaks the spatial symmetry and leads to the anisotropization of
transport coefficients
 \cite{Huang:2009ue,Huang:2011dc,Critelli:2014kra,Finazzo:2016mhm,Hernandez:2017mch}.
Their values depend on the strength of the magnetic field.

The transport coefficients are important physical quantities characterizing the features of QGP and reflecting the nature
of interactions between quarks and gluons. In the last decade, dissipative hydrodynamic models 
\cite{Shen:2011zc,Romatschke:2007mq,Song:2007fn,Bozek:2012qs,Roy:2012jb,Niemi:2012ry,Heinz:2011kt,Schenke:2011bn,Romatschke:2017ejr} 
have played a very important role in extracting the shear and bulk viscosity of QGP from the flow measurements 
\cite{Schenke:2010rr,Ackermann:2000tr,Adler:2003kt,Adams:2003am,Back:2004zg,Alver:2006wh,Afanasiev:2009wq,Aamodt:2010pa,Xu:2007jv}. 
Now it is necessary to develop models based on relativistic magneto-hydrodynamics (MHD) 
\cite{Groot:1972,Rezzolla:2013}, in order to study QGP dynamics in the presence of magnetic field 
\cite{Gursoy:2014aka,Zakharov:2014dia,Roy:2015kma,Pu:2016ayh,Pu:2016bxy,Pang:2016yuh,Inghirami:2016iru,Das:2017qfi,Greif:2017irh,Stewart:2017zsu,Moghaddam:2017myy,Mohanty:2018eja}. 

How large the magnetic effect (also the chiral magnetic effect 
\cite{Deng:2012pc,Tuchin:2014iua,Zakharov:2014dia,Kharzeev:2015znc}) is, depends on the strength of the magnetic
field. The magnetic field in the early stage of a noncentral heavy-ion collision stems from the pass of two moving ions
and its life time is very short. However, the rapid decrease of this external magnetic field will lead to an electromagnetic
induction of QGP, so that the total magnetic field can last longer, if the QGP is a good conductor with a large electric
conductivity. Therefore, the value of the electric conductivity of the QGP is essential for the possibility of observing
magnetic effects in  heavy-ion collisions.

According to the Green-Kubo relation \cite{Green1954,Kubo:1957mj}, the transport coefficients are related to the
correlation functions of the corresponding tensor or flux.
In Refs. \cite{Huang:2011dc,Finazzo:2016mhm,Hernandez:2017mch} Kubo formulas for anisotropic transport
coefficients in the presence of a magnetic field are derived with different methods. The calculation of the correlation
functions of fluctuating tensor or flux can be realized in kinetic transport models. In this work, we employ the Boltzmann
Approach of MultiParton Scattering (BAMPS) \cite{Xu:2004mz}, which solves the
Boltzmann equation for systems of on-shell particles. In the early studies (in the absence of a magnetic field), 
BAMPS has been used to calculate the shear viscosity of a pQCD-based gluon gas \cite{Xu:2007ns,Wesp:2011yy}
and of QGP \cite{Uphoff:2014cba}, the electric conductivity of QGP \cite{Greif:2014oia} and recently the shear viscosity
of ultrarelativistic Boson systems in the presence of a Bose-Einstein condensation \cite{Chen:2018mwr}.

For a simple case such like a one-component system with massless Boltzmann particles undergoing isotropic binary
elastic collisions, the anisotropic transport coefficients in a magnetic field can be calculated analytically by using Grad's 
approximations \cite{Denicol:2018rbw}.  In this work we consider the same particle system as in \cite{Denicol:2018rbw},
calculate the anisotropic transport coefficients with BAMPS via the Kubo formulas given in 
\cite{Huang:2011dc,Finazzo:2016mhm,Hernandez:2017mch}, and compare the results with those obtained in 
\cite{Denicol:2018rbw}. An agreement of both results will confirm the general use of the derived Kubo formulas for
calculating the anisotropic transport coefficients of QGP in a magnetic field.

The paper is organized as follows: In Sec.~\ref{sec:derivation} we briefly review the equations of
magneto-hydrodynamics and give the Kubo formulas for the corresponding transport coefficients.
In Sec.~\ref{sec:BAMPS} we introduce the parton cascade BAMPS and numerical implementations. Subsequently,
in Sec.~\ref{sec:NumResults} we show our numerical results including the influence of the magnetic field on the time
evolution of corresponding correlation functions and the values of shear viscosity and electric conductivity coefficients
for an one-component system of ultrarelativistic Boltzmann particles with isotropic binary scatterings.
Finally, we give a conclusion in Sec.~\ref{sec:conclusion}.   

We adopt natural units, $\hbar=c=k_B=1$. The metric tensor is chosen to be $g^{\mu\nu}=diag(+,-,-,-)$.

\section{Anisotropic transport coefficients and Kubo formulas}
\label{sec:derivation}
The dynamics of relativistic fluids in external magnetic field has been studied in Ref.\cite{Huang:2011dc}. The authors
have found that due to the breaking of the spatial symmetry in the presence of a magnetic field, the dissipative
functions contain anisotropic transport coefficients, namely, two bulk viscosity, five shear viscosity and three electric
conductivity coefficients.  At first we briefly summarize the results of Ref.\cite{Huang:2011dc}. The same results can
also be found in Ref.\cite{Hattori:2017qih}. 
   
For a charged particle system, the basic equations of MHD consist of the conservation laws of energy, momentum,
and electric charge, and the constitutive equations for the energy-momentum tensor and the electric current. 
The conservation laws can be expressed as
\begin{eqnarray}
\label{jmu_expr}
\partial_{\mu} j^{\mu} &=& 0  \,, \\ 
\label{Tmunu_expr}
\partial_{\mu} T^{\mu \nu} &=& F^{\nu \mu}j_{\mu} \,,
\end{eqnarray}
where $F^{\mu \nu}$ is the electromagnetic field-strength tensor. The electric field is neglected in 
\cite{Huang:2011dc,Finazzo:2016mhm} assuming that the electric field is much smaller than the magnetic field,
which is a good approximation for QGP produced in heavy-ion collisions for instance.  The constitutive equations
in the Landau frame read \cite{Huang:2011dc,Hattori:2017qih} 
\begin{eqnarray}
\label{jmu}
j^{\mu} &=& qnu^{\mu}+ \mathcal{J}^{\mu} \,, \\ 
\label{Tmunu}
T^{\mu \nu} &=& \varepsilon u^{\mu}u^{\nu}-{P}_{\perp}\Xi^{\mu\nu}+{P}_{\parallel}b^{\mu}b^{\nu}+\mathcal{T}^{\mu\nu}\,,
\end{eqnarray}
where $u^{\mu}$ is the fluid 4-velocity normalized to $u^2=1$ and $b^\mu \equiv B^\mu /B$ with 
$B^\mu\equiv\epsilon^{\mu\nu\alpha\beta}F_{\nu\alpha}u_{\beta}/2$ and $B\equiv\sqrt{-B^\mu B_\mu}$. 
$q$ is the particle charge.  $\varepsilon$ and $n$ denote the energy and particle number density, respectively.
The tensor which projects onto the three-dimensional
space orthogonal to the flow velocity $u^\mu$ is defined as $\Delta^{\mu\nu}\equiv g^{\mu\nu}-u^\mu u^\nu$. 
$\Xi^{\mu\nu}\equiv\Delta^{\mu\nu}+b^{\mu}b^{\nu}$ is the tensor, which projects onto the two-dimensional space
orthogonal to both the flow velocity $u^\mu$ and the direction of magnetic field.
$P_{\parallel}$ and $P_{\perp}$ are defined as $P_{\parallel}\equiv  b_\mu b_\nu T^{\mu\nu}$ and
$P_{\perp}\equiv  -\Xi_{\mu\nu} T^{\mu\nu}/2$. The dissipative terms in Eqs.~(\ref{jmu}) and (\ref{Tmunu})
can be obtained by the derivative expansion to the leading order and have the form in terms
of viscosity and electric conductivity coefficients,
\begin{eqnarray}
\label{jmu_coeff}
\mathcal{J}^{\mu} &=& T(\kappa_\perp\Xi^{\mu\nu}\nabla_\nu\alpha - \kappa_\parallel b^\mu b^\nu \nabla_\nu\alpha -\kappa_\times b^{\mu\nu}\nabla_\nu\alpha) \,, \\ 
\label{Tmunu_coeff}
\mathcal{T}^{\mu \nu} &=& \frac{3}{2}\zeta_{\perp}\Xi^{\mu\nu}\phi +3\zeta_\parallel b^\mu b^\nu \psi +2\eta_0 (w^{\mu\nu}-\frac{1}{3}\Delta^{\mu\nu}\theta) \nonumber \\
&&+\eta_1 (\Delta^{\mu\nu}-\frac{3}{2}\Xi^{\mu\nu})(\theta-\frac{3}{2}\phi) \nonumber \\
&& -2\eta_2(b^\mu \Xi^{\nu\alpha}b^\beta+b^\nu \Xi^{\mu\alpha}b^\beta)w_{\alpha\beta} \nonumber \\
&&-2\eta_3(\Xi^{\mu\alpha}b^{\nu\beta}+\Xi^{\nu\alpha}b^{\mu\beta})w_{\alpha\beta} \nonumber \\
&& +2\eta_4(b^{\mu\alpha}b^\nu b^\beta + b^{\nu\alpha}b^\mu b^\beta)w_{\alpha\beta} \,,
\end{eqnarray}
where $\alpha\equiv\beta\mu$ ($\mu$ is the chemical potential), $b^{\mu\nu}\equiv\epsilon^{\mu\nu\alpha\beta}b_\alpha u_\beta$,
$w^{\mu\nu}\equiv (\nabla^\mu u^\nu + \nabla^\nu u^\mu)/2$, $\phi\equiv\Xi^{\mu\nu}w_{\mu\nu}$, 
$\psi\equiv b^\mu b^\nu w_{\mu\nu}$, $\theta\equiv\partial^\mu u_\mu$, with $\nabla_\mu\equiv\triangle_{\mu\nu}\partial^\nu$.
The combining coefficients are identified as five shear viscosity ($\eta_0, \cdots, \eta_4$), two bulk viscosity 
($\zeta_\perp$, $\zeta_\parallel$), and three electric conductivity coefficients 
($\kappa_\perp$, $\kappa_\parallel$, $\kappa_\times$).  

We now summarize Kubo formulas for the anisotropic transport coefficients, which are given in 
Refs.\cite{Huang:2011dc},\cite{Finazzo:2016mhm} and \cite{Hernandez:2017mch}. 
$T^{\mu\nu}$ and $j^\mu$ in the formulas given below are
taken at the local rest frame. The authors of Ref.\cite{Huang:2011dc} used Zubarev's
non-equilibrium statistical operator method to relate the anisotropic transport coefficients to correlation functions in
equilibrium. The corresponding Kubo formulas are given by \cite{Huang:2011dc,Hattori:2017qih} 
\begin{eqnarray}
\label{eta0}
\eta_0 &=& \dfrac{\partial}{\partial \omega} {\rm Im} G^R_{T^{12}T^{12}}\vert_{{\bf p}={\bf 0},\omega\rightarrow 0}\,, \\ 
\label{eta1}
\eta_1 &=& -\frac{4}{3}\eta_0-2\dfrac{\partial}{\partial \omega}{\rm Im} G^R_{\widetilde{P}_{\parallel} \widetilde{P}_{\perp}}\vert_{{\bf p}={\bf 0},\omega\rightarrow 0}\,, \\ 
\label{eta2}
\eta_2 &=& -\eta_0+\dfrac{\partial}{\partial \omega} {\rm Im} G^R_{T^{13}T^{13}}\vert_{{\bf p}={\bf 0},\omega\rightarrow 0}\,, \\ 
\label{eta3}
\eta_3 &=& \frac{1}{4} \dfrac{\partial}{\partial \omega} {\rm Im} G^R_{(T^{11}-T^{22})T^{12}}\vert_{{\bf p}={\bf 0},\omega\rightarrow 0}\,, \\ 
\label{eta4}
\eta_4 &=& \dfrac{\partial}{\partial \omega} {\rm Im} G^R_{T^{13}T^{23}}\vert_{{\bf p}={\bf 0},\omega\rightarrow 0}\,, 
\end{eqnarray}
where the retarded Green's function in quantum statistical theory has the form 
$G^R_{AB}\equiv i \theta(x^0)\left\langle \left[ A(x), B(0) \right]  \right\rangle $, and the angular brackets denote
the ensemble average in equilibrium. Some other symbols in the above formulas are defined as 
$\widetilde{P}_{\parallel}\equiv P_{\parallel}-\Theta_{\beta}\varepsilon-\Theta_{\alpha}n$ and
$\widetilde{P}_{\perp}\equiv P_{\perp}-(\Theta_{\beta}+\Phi_{\beta})\varepsilon-(\Theta_{\alpha}+\Phi_{\alpha})n$ 
with $\Theta_\beta\equiv(\dfrac{\partial P}{\partial\varepsilon})_{n,B}$, 
$\Phi_\beta\equiv -B(\dfrac{\partial M}{\partial\varepsilon})_{n,B}$, 
$\Theta_\alpha\equiv(\dfrac{\partial P}{\partial n})_{\varepsilon,B}$, and 
$\Phi_\alpha\equiv -B(\dfrac{\partial M}{\partial n})_{\varepsilon,B}$. 
$M\equiv (\partial P/ \partial B)_{T,\mu}$ is the magnetization. More details can be found in 
\cite{Huang:2011dc,Hattori:2017qih}. The coefficients involving magnetization in the definition of 
$\widetilde{P}_{\parallel}$ and $\widetilde{P}_{\perp}$ vanish for particles without dipole moment or spin,
which is the case we consider in this work. We also note that a sign mistake in the formula
of $\eta_3$ occurred in \cite{Huang:2011dc} has been corrected.

The Kubo formulas of the viscosity coefficients were also given in Refs. \cite{Finazzo:2016mhm,Hernandez:2017mch},
where a variational approach and a derivative method were used, respectively.  Despite the sign and/or factor differences
after unifying the convention for the transport coefficients, the Kubo formulas for the five shear viscosity coefficients
are definitely the same among \cite{Huang:2011dc}, \cite{Finazzo:2016mhm} and \cite{Hernandez:2017mch}. 
Since the bulk viscosity coefficients vanish by considering a massless Boltzmann gas, there is no need to give their
Kubo formulas. 

The Kubo formulas for the shear viscosity coefficients can be expressed in real space-time with a integration form\cite{Zubarev1997,Calzetta2008,Searles2000,Paech:2006st}:
\begin{eqnarray}
\label{Ieta0}
\eta_0 &=& \dfrac{1}{T}\int_0^\infty dt \int_V d^3r \left\langle T^{12}({\bf r}, t)T^{12}(0,0)\right\rangle  \,, \\ 
\label{Ieta1}
\eta_1 &=& -\frac{4}{3}\eta_0-\dfrac{2}{T}\int_0^\infty dt \int_V d^3r \left\langle \widetilde{P}_{\parallel}({\bf r}, t)\widetilde{P}_{\perp}(0,0)\right\rangle \,, \\
\label{Ieta2}
\eta_2 &=& -\eta_0+\dfrac{1}{T}\int_0^\infty dt \int_V d^3r \left\langle T^{13}({\bf r}, t)T^{13}(0,0)\right\rangle \,, \\ 
\label{Ieta3}
\eta_3 &=& \dfrac{1}{4T}\int_0^\infty dt \int_V d^3r \\ \nonumber
&&\times \left\langle (T^{11}({\bf r}, t)-T^{22}({\bf r}, t))T^{12}(0,0)\right\rangle \,, \\ 
\label{Ieta4}
\eta_4 &=& \dfrac{1}{T}\int_0^\infty dt \int_V d^3r \left\langle T^{13}({\bf r}, t)T^{23}(0,0)\right\rangle\,, 
\end{eqnarray}
where $T$ is the temperature and the angular brackets denote the ensemble average. Without loss of generality,
we choose the z-direction as the direction of the magnetic field, $B^\mu=(0,0,0,B_0)$. Since we consider
a homogeneous particle system, the space dependence of the particle flow and the energy-momentum tensor
appeared in the above Kubo formulas can be integrated out directly. 

Without the magnetic field ($B_0=0$), except for $\eta_0$,  which should be equal to the standard isotropic
shear viscosity $\eta$, all other shear viscosity coefficients should vanish. It is obvious for $\eta_0$, $\eta_2$, 
$\eta_3$, and $\eta_4$, since $\left\langle T^{13}({\bf r}, t)T^{13}(0,0)\right\rangle$ is equal to 
$\left\langle T^{12}({\bf r}, t)T^{12}(0,0)\right\rangle$ and both 
$\left\langle (T^{11}({\bf r}, t)-T^{22}({\bf r}, t))T^{12}(0,0)\right\rangle$ and  
$\left\langle T^{13}({\bf r}, t)T^{23}(0,0)\right\rangle$ vanish. For the considered system we have
 $\left\langle \widetilde{P}_{\parallel}({\bf r}, t)\widetilde{P}_{\perp}(0,0)\right\rangle = -\left\langle \mathcal{T}^{33}({\bf r}, t)\mathcal{T}^{33}(0,0)\right\rangle/2$
according to the definitions of $ \widetilde{P}_{\parallel}$ and $\widetilde{P}_{\perp}$. From \cite{Zubarev1997,Wesp:2011yy} 
we realize that $\left\langle \mathcal{T}^{33}({\bf r}, t)\mathcal{T}^{33}(0,0)\right\rangle=4\left\langle \mathcal{T}^{12}({\bf r}, t)\mathcal{T}^{12}(0,0)\right\rangle/3=4\left\langle T^{12}({\bf r}, t)T^{12}(0,0)\right\rangle/3$. Therefore, $\eta_1$ vanishes.

Due to the breaking of spatial symmetry by a longitudinal magnetic field, the usual isotropic electric conductivity
becomes anisotropic. The transverse conductivity is different from the longitudinal one. Charged particles in the
magnetic field experience the Lorentz force, which results in an electric current in the perpendicular direction with
respect to both electric and magnetic fields. The electric conductivity associated with this current is called Hall
conductivity. 

In Ref. \cite{Huang:2011dc}  the electric conductivity coefficients associated with the diffusion (or heat transfer)
are given by the following Kubo formulas:
\begin{eqnarray}
\label{kappa1}
\kappa_{\parallel} &=& \dfrac{\partial}{\partial \omega} {\rm Im} G^R_{G^3G^3}\vert_{{\bf p}={\bf 0},\omega\rightarrow 0}\,, \\ 
\label{kappat}
\kappa_{\perp} &=& \dfrac{\partial}{\partial \omega} {\rm Im} G^R_{G^1G^1}\vert_{{\bf p}={\bf 0},\omega\rightarrow 0}\,, \\ 
\label{kappax}
\kappa_{\times} &=& \dfrac{\partial}{\partial \omega} {\rm Im} G^R_{G^1G^2}\vert_{{\bf p}={\bf 0},\omega\rightarrow 0}\,,
\end{eqnarray}
where $G^i(t)=qT^{0i}/4T-j^i$ with $i=1,2,3$ denoting the space components. 
$\kappa_{\parallel}, \kappa_{\perp}$, and $\kappa_{\times}$ are the longitudinal, transverse, and Hall electric
conductivity, respectively.

Different from the definitions of $\kappa_{\parallel}, \kappa_{\perp}$, and $\kappa_{\times}$,
in Ref.\cite{Hernandez:2017mch} the electric conductivity (resistivity) coefficients are induced by an electric 
and a magnetic field, which are perpendicular. The corresponding Kubo formulas are given by
\begin{eqnarray}
\label{kappa1_hk}
\frac{1}{\omega} {\rm Im} G^R_{j^3j^3}(\omega,{\bf p}={\bf 0}) &=& \sigma_{\parallel}  \,, \\
\label{kappat_hk1}
\frac{1}{\omega} {\rm Im} G^R_{T^{01}T^{01}}(\omega,{\bf p}={\bf 0}) &=& \rho_\perp \frac{\omega^2_0}{B^2_0}  \,, \\
\label{kappax_hk1}
\frac{1}{\omega} {\rm Im} G^R_{T^{01}T^{02}}(\omega,{\bf p}={\bf 0}) &=& -\widetilde{\rho}_\perp \frac{\omega^2_0}{B^2_0}sign(B_0)  \,, 
\end{eqnarray}
where $\sigma_\parallel$ denotes the longitudinal electric conductivity, and $\rho_\perp$, $\widetilde{\rho}_\perp$
denote the transverse and Hall electric resistivity, respectively.  $\omega_0=\varepsilon+P$ is the enthalpy density.
All of these formulas are in the limit ${\bf p} \rightarrow {\bf 0}$ and $\omega\rightarrow 0$. We note that 
$\sigma_\parallel, \rho_\perp$, and $\widetilde{\rho}_\perp$ have no relation to the diffusion (or heat transfer).

We give now the Kubo formulas in real space-time:
\begin{eqnarray}
\label{Ikappa1}
\kappa_{\parallel} &=& \dfrac{1}{T}\int_0^\infty dt \int_V d^3r \left\langle G^3({\bf r}, t)G^3(0,0)\right\rangle  \,, \\
\label{Ikappat}
\kappa_{\perp} &=& \dfrac{1}{T}\int_0^\infty dt \int_V d^3r \left\langle G^1({\bf r}, t)G^1(0,0)\right\rangle  \,, \\
\label{Ikappax}
\kappa_{\times} &=& \dfrac{1}{T}\int_0^\infty dt \int_V d^3r \left\langle G^1({\bf r}, t)G^2(0,0)\right\rangle  \,, \\
\label{Ikappa1_hk}
\sigma_{\parallel} &=& \dfrac{1}{T}\int_0^\infty dt \int_V d^3r \left\langle j^3({\bf r}, t)j^3(0,0)\right\rangle  \,, \\
\label{Ikappat_hk}
\rho_\perp &=& \dfrac{B^2_0}{\omega^2_0 T}\int_0^\infty dt \int_V d^3r \left\langle T^{01}({\bf r}, t)T^{01}(0,0)\right\rangle  \,, \\
\label{Ikappax_hk}
\widetilde{\rho}_\perp &=& -sign(B_0)\dfrac{B^2_0}{\omega^2_0 T}\int_0^\infty dt \int_V d^3r \\ \nonumber
&&\times \left\langle T^{01}({\bf r}, t)T^{02}(0,0)\right\rangle  \,.
\end{eqnarray} 
Differences in numerical results of two kind of electric conductivity coefficients will be shown later in 
Sec. \ref{sec:NumResults}.

Without the magnetic field ($B_0=0$), the longitudinal and transverse electric conductivity become
equal, while the Hall electric conductivity (or resistivity) is meaningless. From the above Kubo formulas, it is
obvious that $\kappa_{\parallel}=\kappa_{\perp}$ and $1/\rho_\perp$ is infinite. We will show in Sec.  \ref{sec:NumResults}
that $\sigma_\parallel$ is also infinite.

\section{The parton cascade BAMPS and numerical implementations}\label{sec:BAMPS}
The time correlation functions in Eqs.(\ref{Ieta0})-(\ref{Ieta4}) and Eqs.(\ref{Ikappa1})-(\ref{Ikappax_hk}) are evaluated
numerically for the considered particle system in a static box with periodic boundary conditions. Initially, particles are
distributed homogeneously in coordinate space and thermally in momentum space. The space-time evolution of 
particles is calculated by employing the parton cascade BAMPS \cite{Xu:2004mz}. Coupled to an external 
electromagnetic field, the Boltzmann equation \cite{Cercignani2002,deGroot} has the form 
\begin{equation}
p^{\mu} \partial_{\mu} f(x,p) + qF^{\mu\nu}p_\nu \dfrac{\partial}{\partial p^\mu} f(x,p) =C[f(x,p)] \,,
\label{Boltzmann_eq}
\end{equation}
where $f(x,p)$ is the one-particle phase-space distribution function, $C[f(x,p)]$ denotes the collision term. 
Since we restrict ourselves to a single-component gas of particles carrying no dipole moment or spin, 
the magnetic field $F^{\mu\nu}$ involves only a Lorentz force, which changes the momenta of charged particles. 
The microscopic interaction processes among particles are simulated via Monte Carlo techniques based on 
the stochastic interpretation of transition rates \cite{Xu:2004mz}. In order to improve the numerical accuracy,
the test particle method \cite{Xu:2004mz} is introduced. The particle number is artificially
increased by a factor of $N_{test}$, while the interaction cross section is reduced by the same factor simultaneously.
Thus, the physical evolution of the particle system is not influenced by this implementation. The collision probability
for binary elastic scattering in a spatial cell of a volume of $\Delta V$ and within a time step $\Delta t$ is
\begin{equation}
P_{22}=v_{rel} \frac{\sigma_{22}}{N_{test}} \frac{\Delta t}{\Delta V} \,.
\end{equation}
$v_{rel}=s/(2E_1E_2)$ is the relative velocity of the incoming particles with energy $E_1$ and $E_2$, 
$s$ is the invariant mass, and $\sigma_{22}$ is the total cross section of elastic binary scatterings. We consider
the magnetic field $\vec{B}$ to be constant and homogeneous, pointing in $z$ direction. Thus, the Lorentz force, 
$\vec{F}_L=q\vec{v}\times \vec{B}$, will change the directions (while not the magnitude) of particles' transverse
momenta for every computational time step $\Delta t$. Between the collisions the particles will move in a circle
in the transverse plane, while they propagate via free streaming in the z direction. 

According to the physical definition, the electric current $j^\mu$ and energy-momentum tensor $T^{\mu\nu}$ are
calculated as 
\begin{eqnarray}
\label{eq:nmu_BAMPS}
{j}^{\mu}(t) &=& \frac{q}{V N_{test}}\sum\limits_{i=1}^{N}\frac{p_{i}^{\mu
}}{E_{i}} \,, \\
\label{eq:tmunu_BAMPS}
{T}^{\mu \nu }(t) &=& \frac{1}{V N_{test}}\sum\limits_{i=1}^{N}\frac{p_{i}^{\mu
}p_{i}^{\nu }}{E_{i}} \,, 
\end{eqnarray}
where the sum is running over all the $N$ test particles in the box at time $t$ and $V$ is the volume of the box. 
The correlation is calculated at discrete and equally distributed time steps $t_l={t_0,t_1...,t_K}$ by time average
in the limit $t_K\rightarrow\infty$, 
\begin{equation}
C(t_l)= \frac{1}{s_{max}} \sum^{s_{max}}_{s=0} A(t_s)B(t_s+t_l)\,, \ \ s_{max}=K-l.
\end{equation}
$A(t)$ and $B(t)$ represent the component of the particle flow or energy-momentum tensor.
The ensemble average is realized by $N_{run}$ individual initialization.
In the presence of a magnetic field, some of the correlation functions oscillate instead of an exponential
decrease (as we will show in the next section). Therefore, we have to evolve the systems to a much longer
time until the correlations become small.

\section{Numerical results}
\label{sec:NumResults}

\begin{figure*}[t]
\includegraphics[width=8cm,height=6cm]{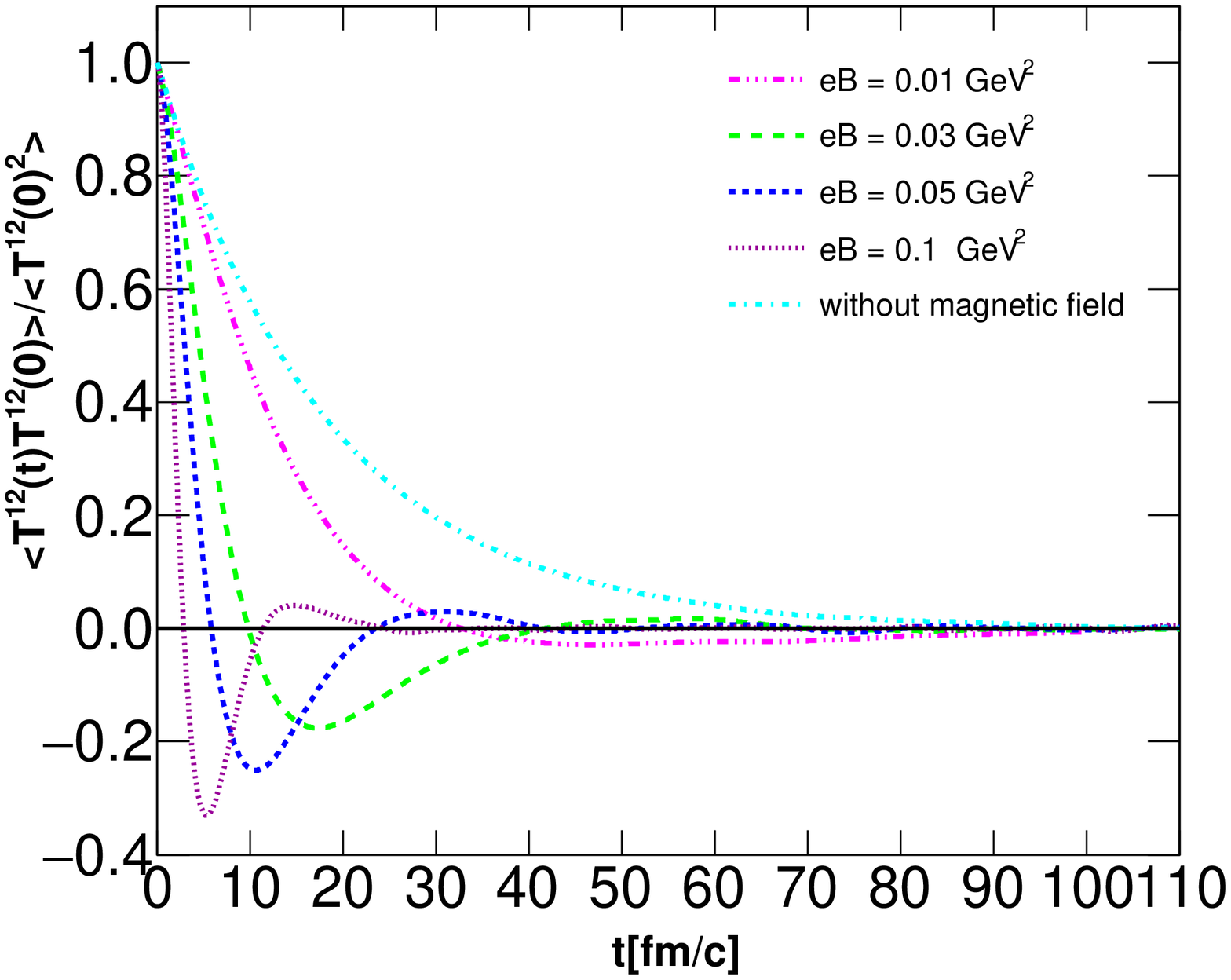}
\includegraphics[width=8cm,height=6cm]{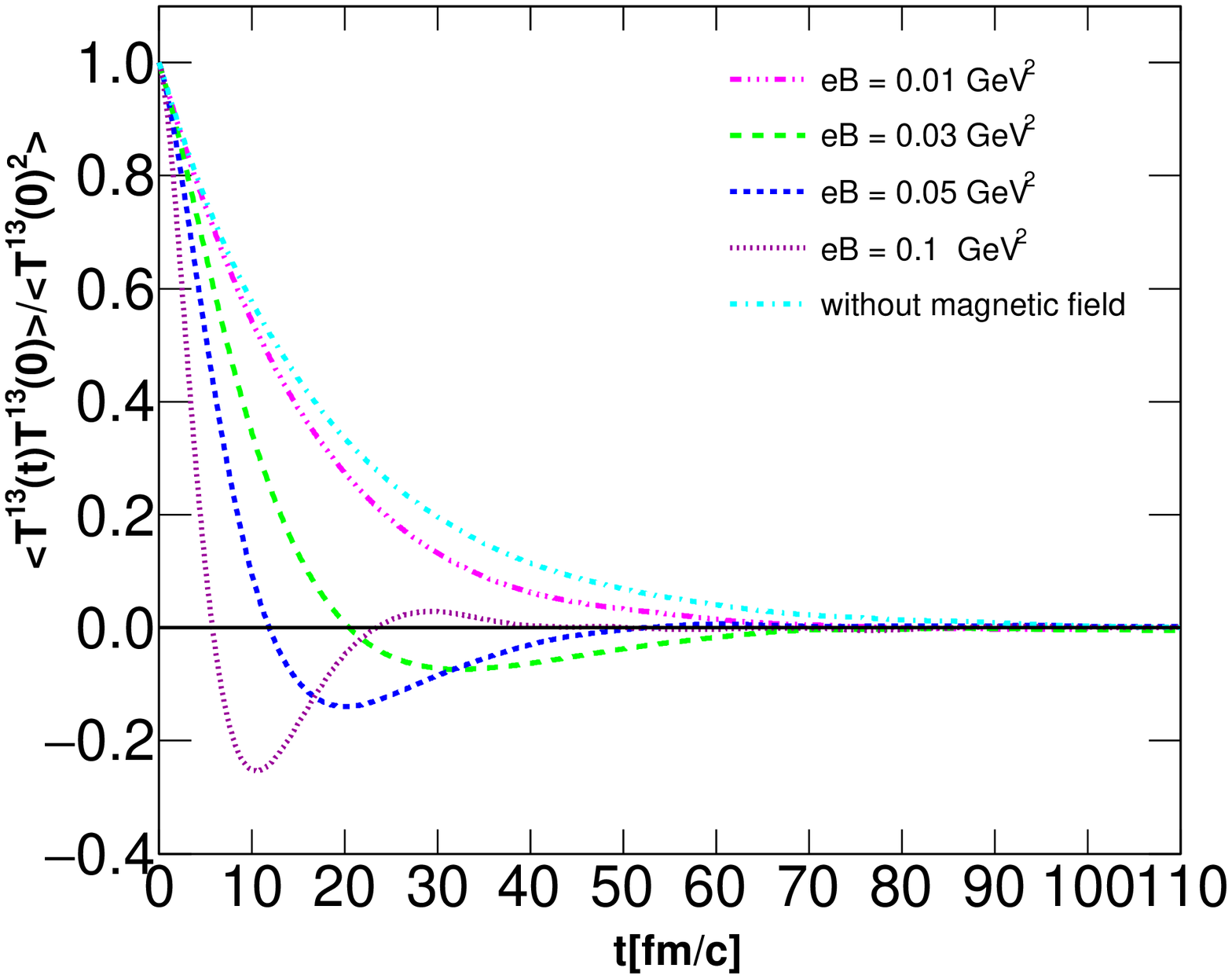}
\caption{\label{fig:correlation0} Time evolution of $\left\langle T^{12}(t)T^{12}(0)\right\rangle$ and $\left\langle T^{13}(t)T^{13}(0)\right\rangle$ with various magnetic field strengths. The results are normalized by their initial values.}
\end{figure*}

\begin{figure}
\includegraphics[width=8cm,height=6cm]{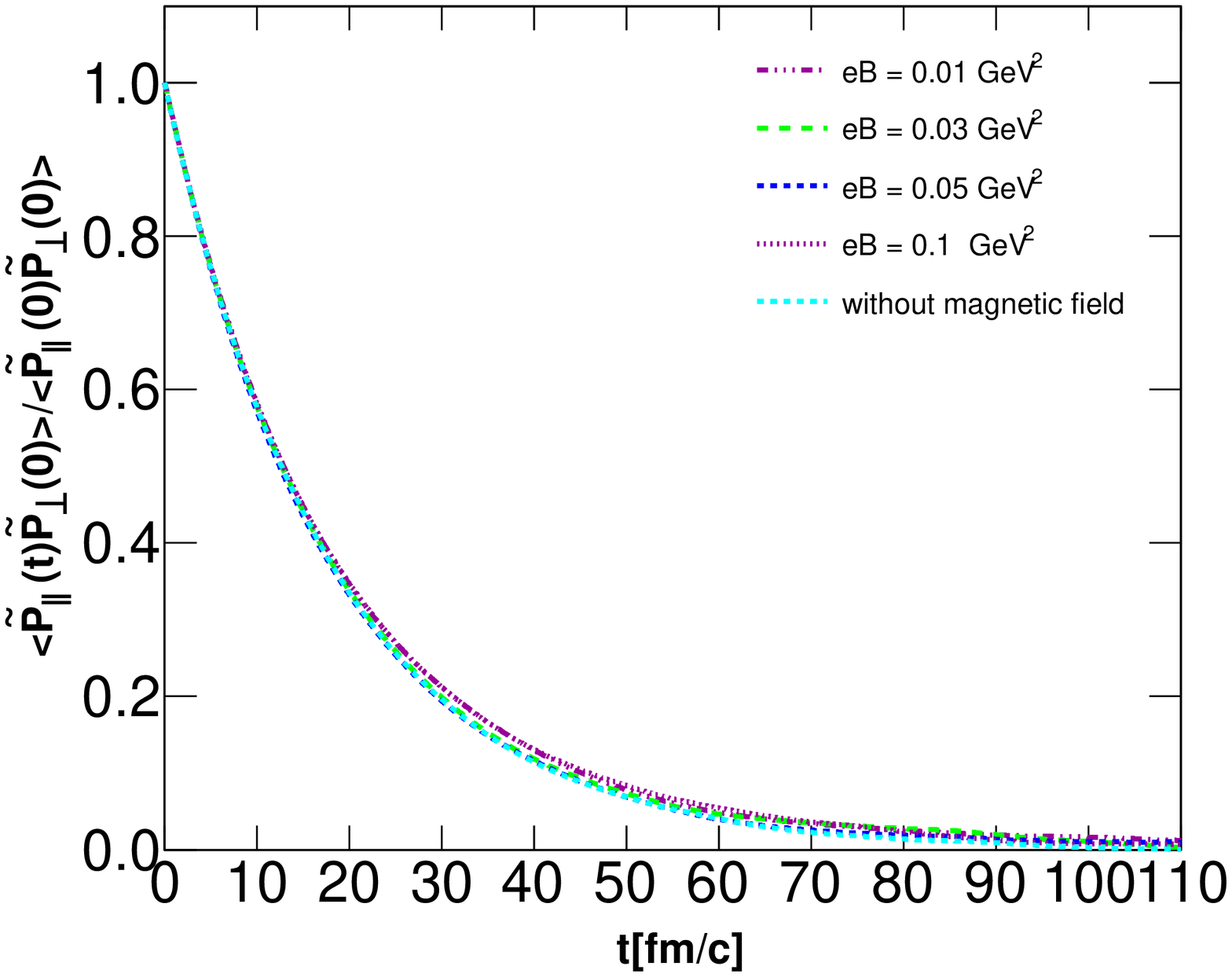}
\caption{\label{fig:correlation1} Same as Fig. \ref{fig:correlation0}, but for $\left\langle \widetilde{P}_{\parallel}(t) \widetilde{P}_{\perp}(0) \right\rangle$.}
\end{figure}

\begin{figure*}[t]
\includegraphics[width=8cm,height=6cm]{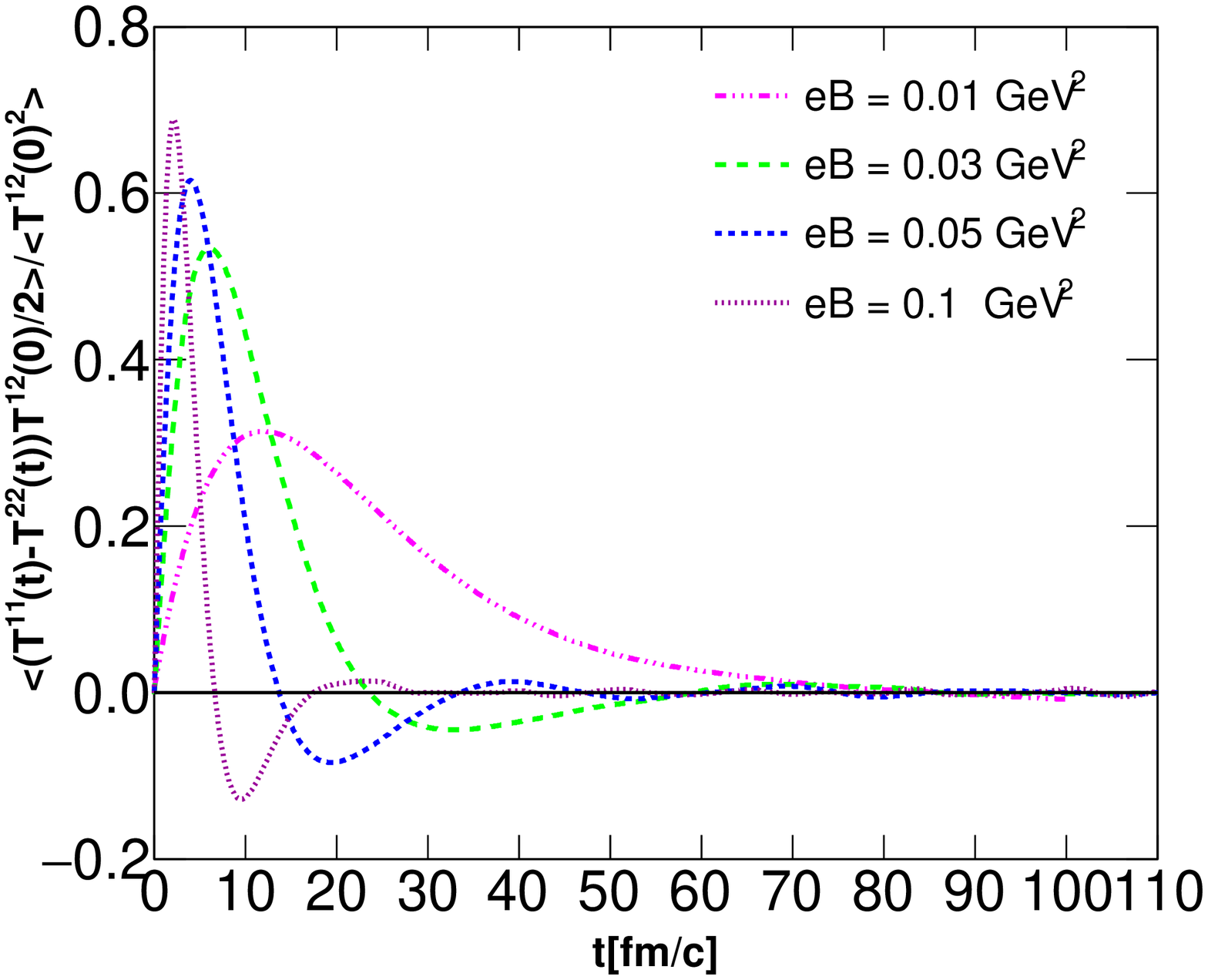}
\includegraphics[width=8cm,height=6cm]{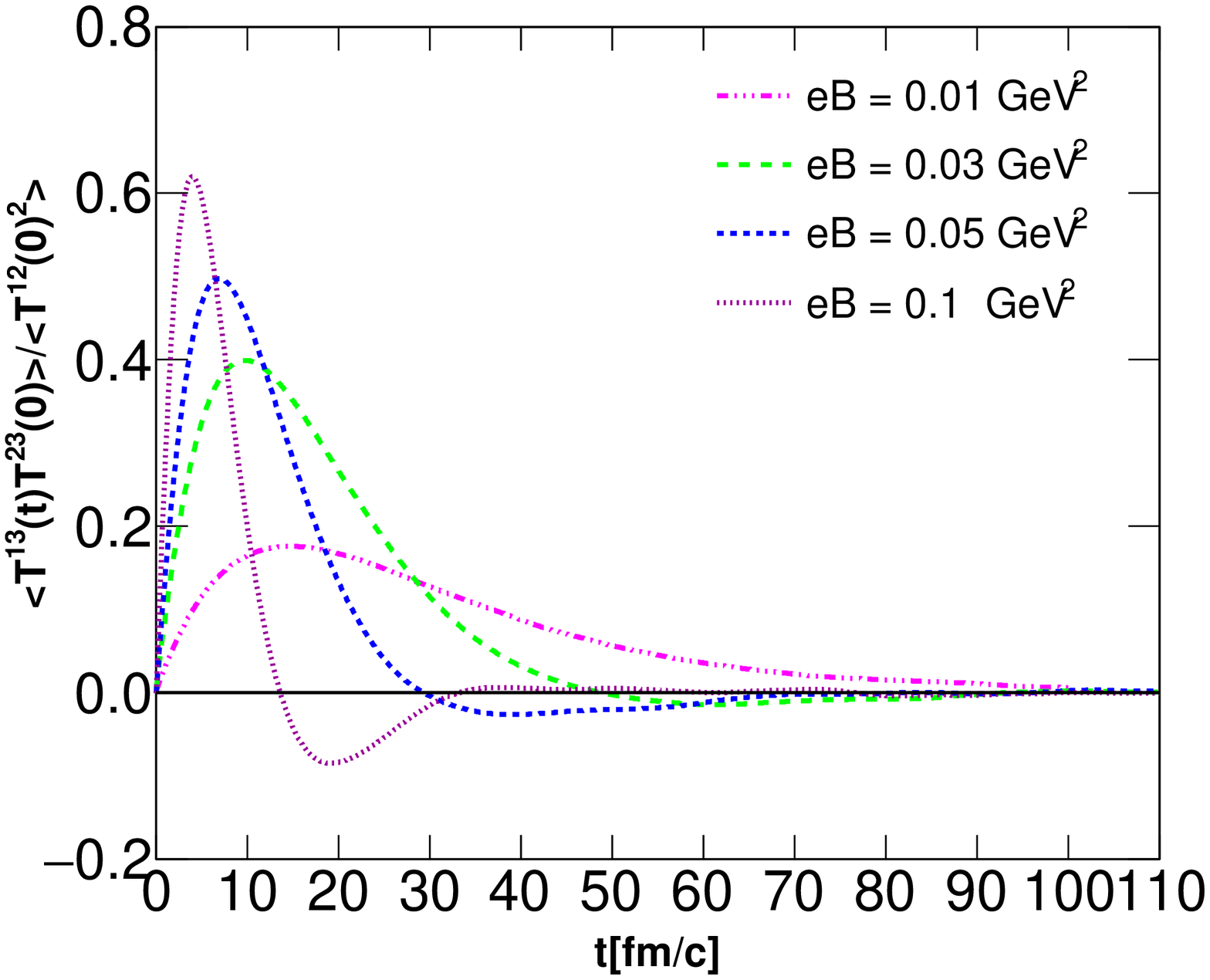}
\caption{\label{fig:correlation3}  Time evolution of $\left\langle (T^{11}(t)-T^{22}(t))T^{12}(0)\right\rangle/2$ and 
$\left\langle (T^{13}(t))T^{23}(0)\right\rangle$ with various magnetic field strengths. The results are normalized by 
$\left\langle (T^{12}(0))^2 \right\rangle$. The results without the magnetic field are zero.}
\end{figure*}

We consider a system of massless Boltzmann particles with a positive charge $e$. Initially particles are
sampled in momentum space according to the Boltzmann distribution $f(x,p)=e^{-E/T}$ with a temperature
of $T = 400 \mbox{ MeV}$. Since the magnetic fields in noncentral Au-Au collisions at RHIC can reach
$eB \sim m^2_\pi = (0.443 \mbox{ GeV})^2$ \cite{Kharzeev:2007jp}, we choose in this work
$eB=0, 0.01, 0.03, 0.05, 0.1 \mbox{ GeV}^2$. We neglect the Landau quantization of the particles' 
cyclotron motion and assume that particles carry no dipole moment or spin, so that the gas has
vanishing magnetization and polarization. We consider binary elastic collisions only. 
The total cross section is set to be a constant value ($\sigma_{22}=1 \mbox{ mb}$), and the particles
scatter isotropically.  

In the following we show the results of shear viscosity and electric conductivity coefficients
in the presence of a magnetic field, respectively.
 
\subsection{Shear viscosity coefficients}
\label{sec:level1}
The time evolution of correlation functions $\left\langle T^{12}(t)T^{12}(0)\right\rangle$ and 
$\left\langle T^{13}(t)T^{13}(0)\right\rangle$, which determine the shear viscosity coefficients $\eta_0$ and $\eta_2$,
are shown in Fig.~\ref{fig:correlation0}. We can see that the correlation functions behave quite different from those
without magnetic field. The correlation functions decrease no longer exponentially. The presence of the magnetic field
induces the oscillations of the correlation functions, because the Lorentz force changes the sign of $T^{12}$ and $T^{13}$.
Compared with $\left\langle T^{12}(t)T^{12}(0)\right\rangle$, the oscillation frequencies of 
$\left\langle T^{13}(t)T^{13}(0)\right\rangle$ are smaller, since the momenta in z-direction are not influenced
by the Lorentz force. 

The correlation function $\left\langle \widetilde{P}_{\parallel}(t) \widetilde{P}_{\perp}(0) \right\rangle$,
which determines the shear viscosity coefficient $\eta_1$, are shown in Fig.~\ref{fig:correlation1}.
For the system we have considered in this work, we have $\widetilde{P}_{\parallel}=T^{33}-T^{00}/3$
and $\widetilde{P}_{\perp}=(T^{11}+T^{22})/2-T^{00}/3=T^{00}/6-T^{33}/2$. From Fig.~\ref{fig:correlation1}
we see that the magnetic field has no influence on this correlation function despite the small deviation
due to numerical fluctuations. This is because the correlation function only involve particles' energy and
momentum in the z-direction, both of them are not affected by the magnetic field.

It is obvious that the correlation functions $\left\langle (T^{11}(t)-T^{22}(t))T^{12}(0)\right\rangle$ and
$\left\langle (T^{13}(t))T^{23}(0)\right\rangle$ will vanish, if there is no magnetic field. With the magnetic field
the two correlation functions, corresponding to $\eta_3$ and $\eta_4$ respectively, oscillate due to the same
reason for $\left\langle T^{12}(t)T^{12}(0)\right\rangle$ and $\left\langle T^{13}(t)T^{13}(0)\right\rangle$,
as seen in Fig.~\ref{fig:correlation3}. The two correlation functions behave similarly. The stronger the magnitude
of the magnetic field, the earlier the maximum value is reached and the larger is the value. 

\begin{figure}
\includegraphics[width=8cm,height=6cm]{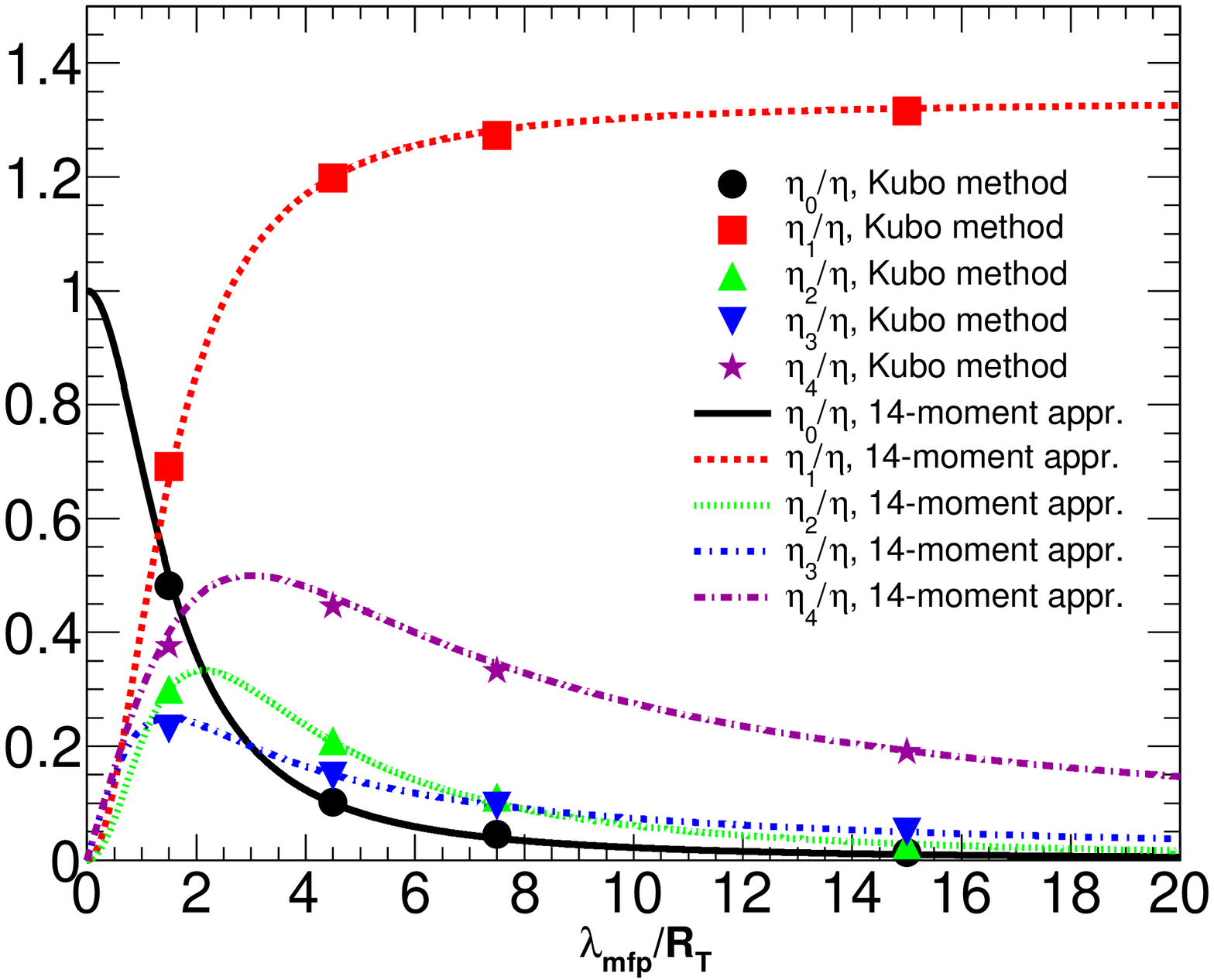}
\caption{\label{fig:eta} The magnetic field dependence of the shear viscosity coefficients scaled by 
the standard isotropic shear viscosity $\eta$. The analytical results from Eq.~(\ref{eta0_moment}) are
shown by the curves for comparisons.}
\end{figure}

\begin{figure}
\includegraphics[width=8cm,height=6cm]{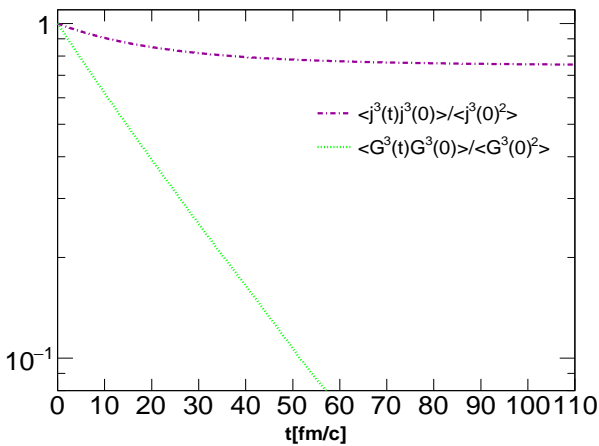}
\caption{\label{fig:gzgz} Time evolution of $\left\langle G^3(t)G^3(0)\right\rangle$ and $\left\langle j^3(t)j^3(0)\right\rangle$.
The results are normalized by their initial values.}
\end{figure}

\begin{figure*}[t]
\includegraphics[width=8cm,height=6cm]{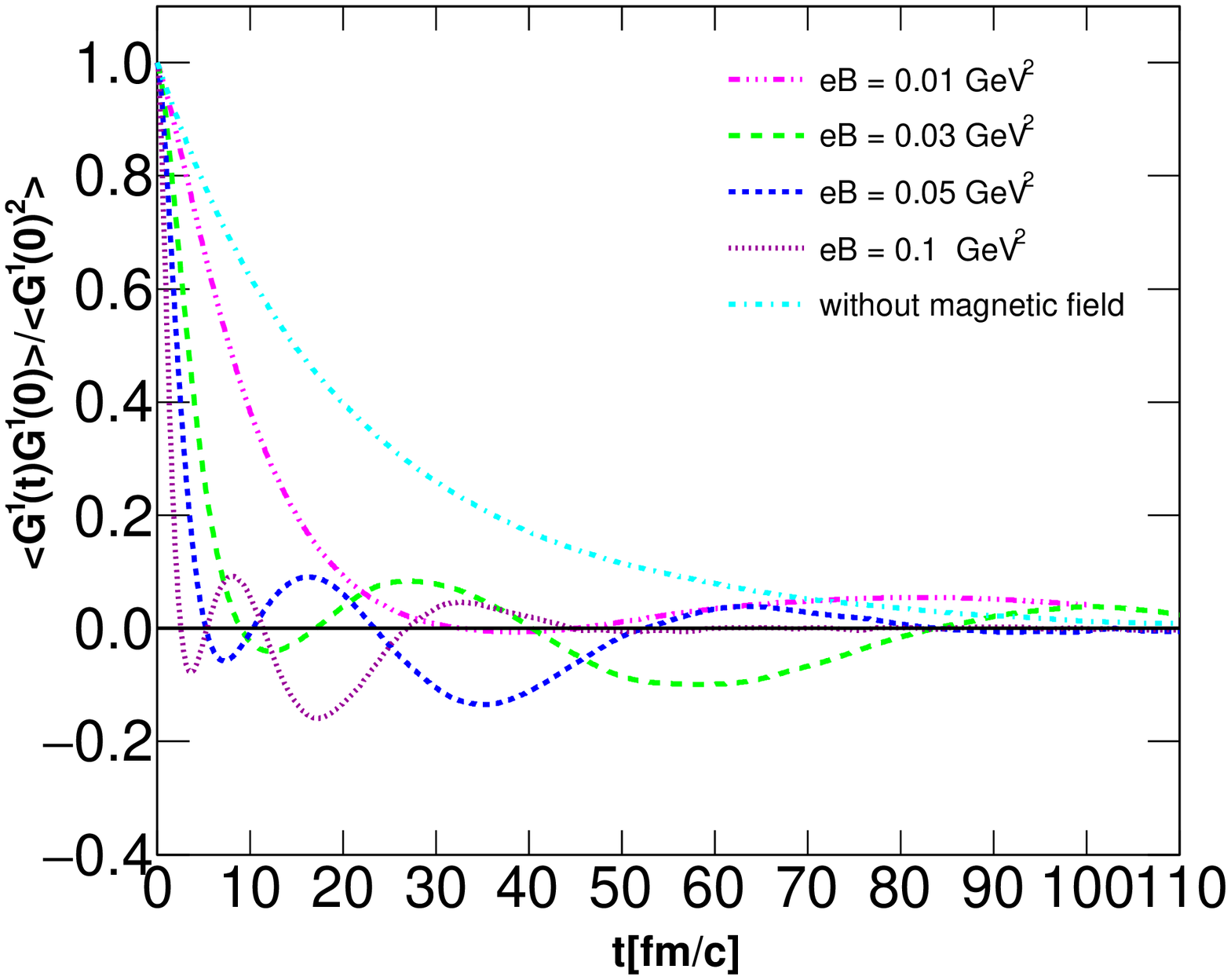}
\includegraphics[width=8cm,height=6cm]{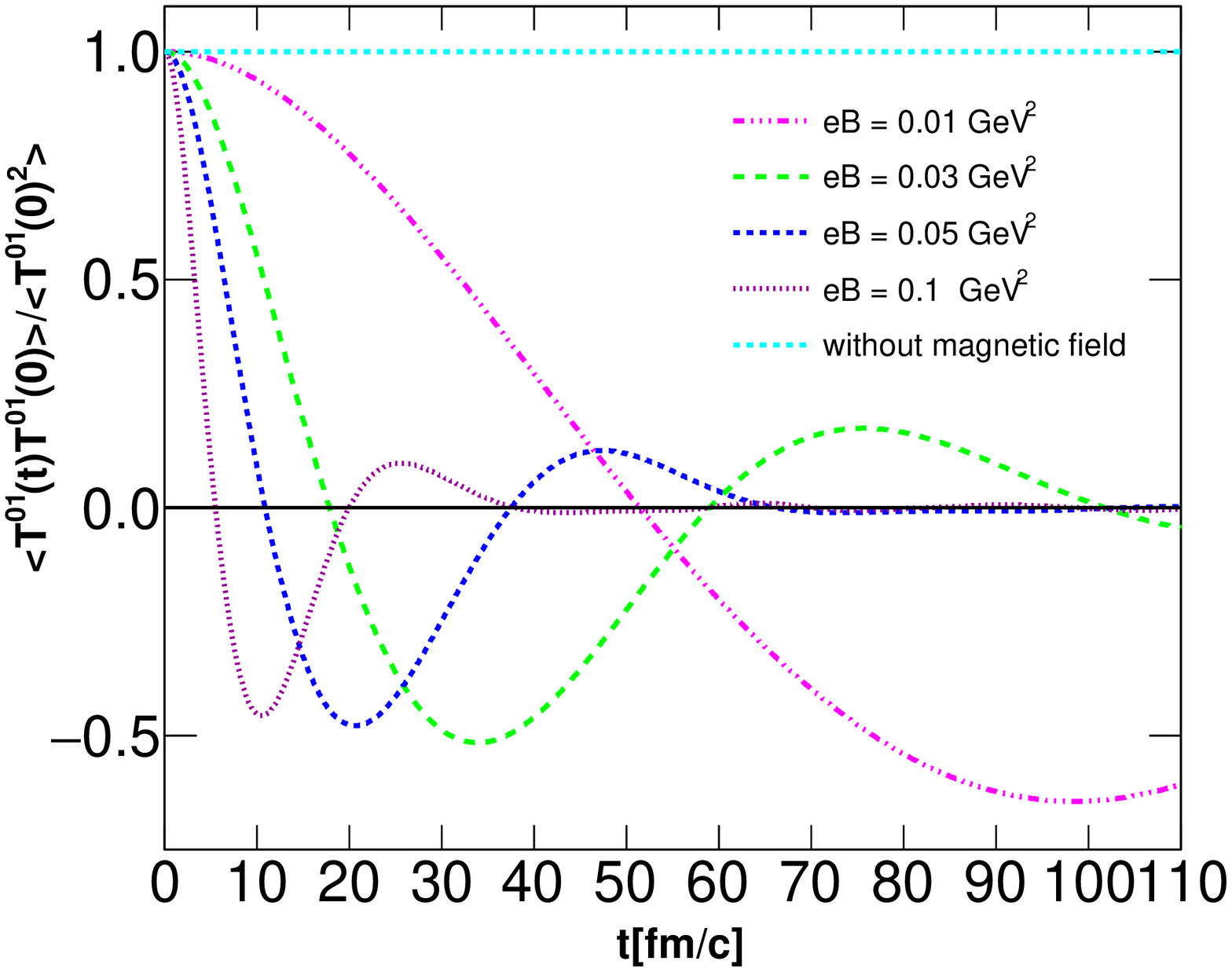}
\caption{\label{fig:gxgx} Same as Fig. \ref{fig:correlation0}, but for $\left\langle G^1(t)G^1(0)\right\rangle$
and $\left\langle T^{01}(t)T^{01}(0)\right\rangle$.}
\end{figure*}

\begin{figure*}[t]
\includegraphics[width=8cm,height=6cm]{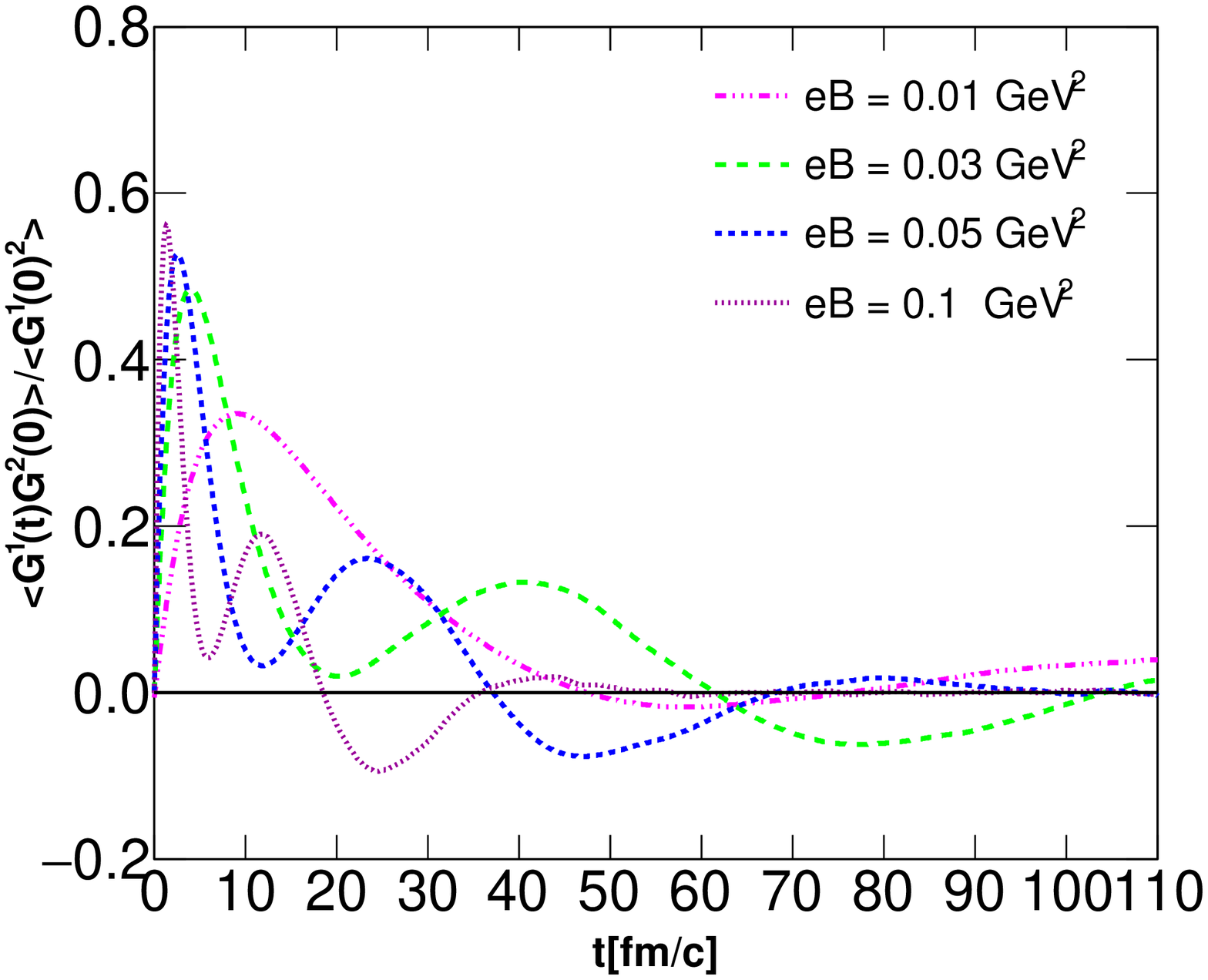}
\includegraphics[width=8cm,height=6cm]{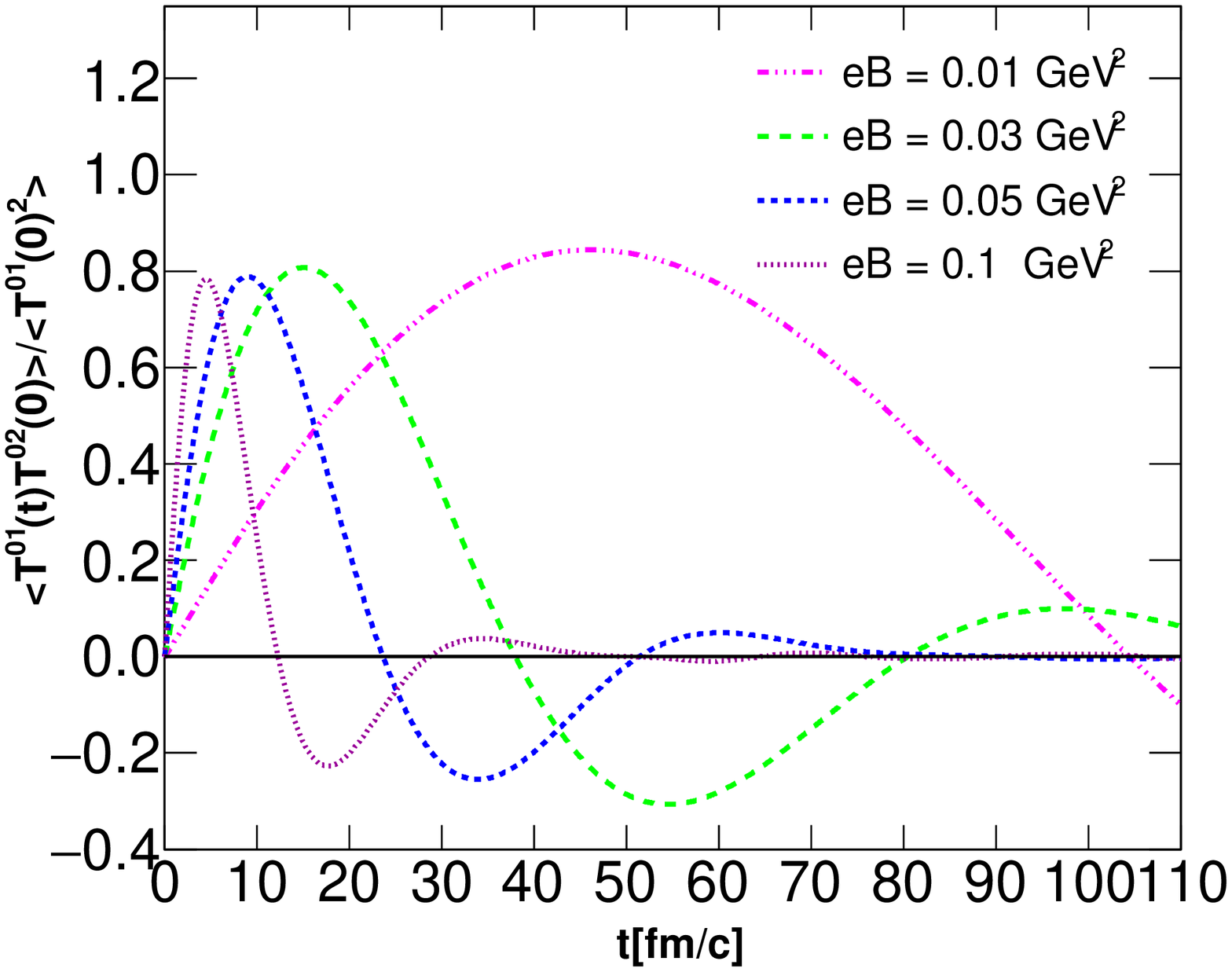}
\caption{\label{fig:gxgy} Time evolution of $\left\langle G^1(t)G^2(0)\right\rangle$ and 
$\left\langle T^{01}(t)T^{02}(0)\right\rangle$ with various magnetic field strengths. The results are normalized by
$\left\langle (G^1(0))^2 \right\rangle$ and $\left\langle (T^{01}(0))^2 \right\rangle$, respectively.
The results without the magnetic field are zero.}
\end{figure*}

\begin{figure}
\includegraphics[width=8cm,height=12cm]{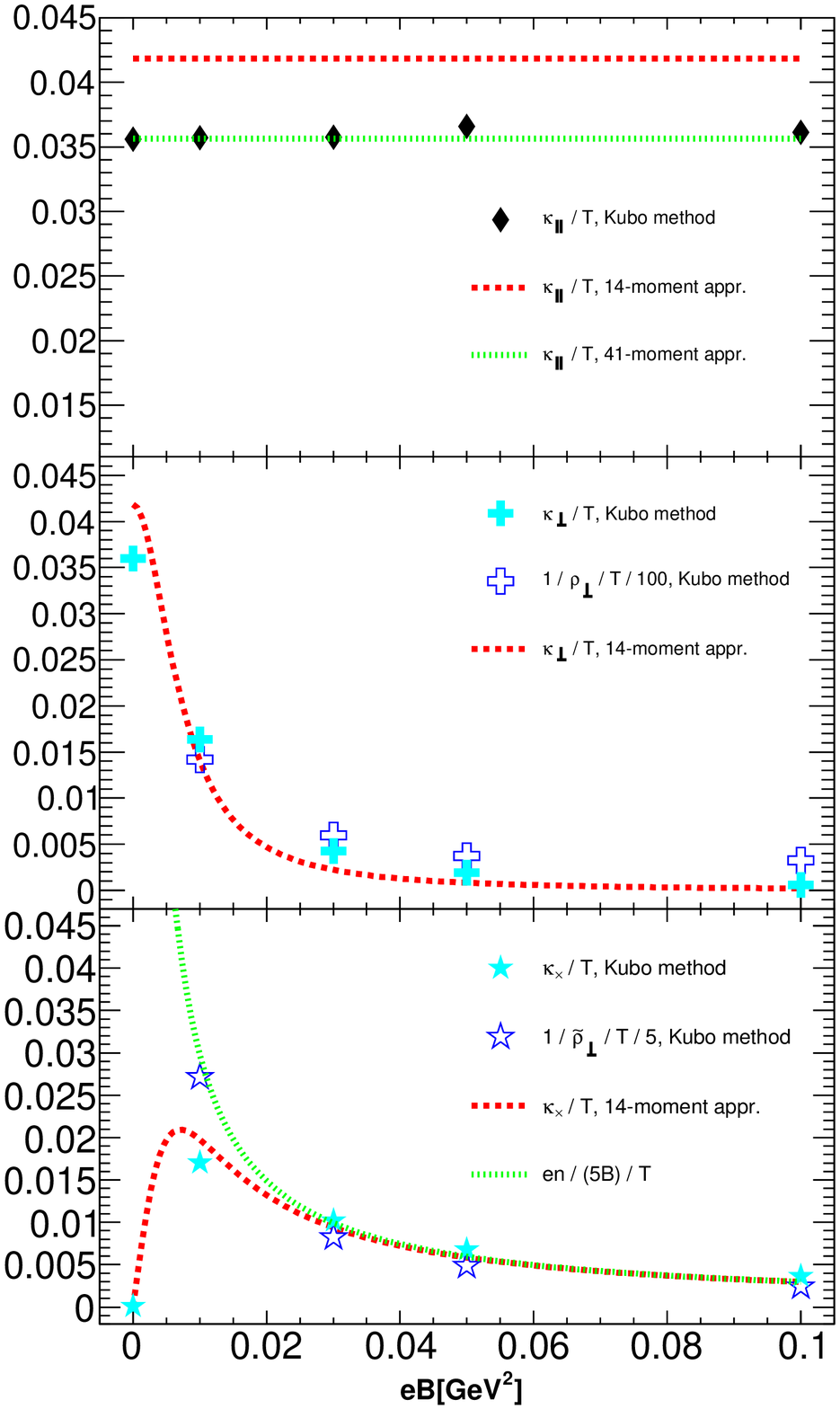}
\caption{\label{fig:kappa} The magnetic field dependence of the electric conductivity coefficients
scaled by the temperature. The analytical results from the $14$-moment approximation 
[see Eq.~(\ref{kappa_moment})] and from the $41$-moment approximation \cite{Denicol:2012cn} are
shown by the curves for comparisons.}
\end{figure}

Figure \ref{fig:eta} shows the five anisotropic shear viscosity coefficients for various values of
$\xi_B=\lambda_{mfp}/R_T$, where the mean free path $\lambda_{mfp}=1/(n\sigma_{22})$ is fixed and
$R_T=T/(eB)$ varies by varying $eB$. $R_T$ denotes the Larmor radius of a particle with the transverse
momentum being equal to the temperature. The five shear viscosity coefficients are normalized by the standard
isotropic shear viscosity $\eta=4\lambda_{mfp}P/3$ obtained without the magnetic field. We see that except for
$\eta_1$, which is almost constant at large magnetic field, all the other shear viscosity coefficients are decreasing. 

The five anisotropic shear viscosity coefficients of a massless Boltzmann gas undergoing binary isotropic
elastic collisions in a magnetic field have been derived analytically in Ref.\cite{Denicol:2018rbw} by using
the method of 14-moment Grad's approximation. We list the results below:
\begin{eqnarray}
\label{eta0_moment}
\eta_0 &=& \frac{12\lambda_{mfp}P}{9+4\xi^2_B} \,,\,\, \eta_1 = \frac{64}{9} \frac{\xi_B^2\lambda_{mfp}P}{9+4\xi_B^2} \,, 
\nonumber  \\ 
\eta_2 &=& \frac{36\xi_B^2\lambda_{mfp}P}{[9+4\xi_B^2][9+\xi_B^2]} \,,\,\, \eta_3 = \frac{4\xi_B\lambda_{mfp}P}{9+4\xi^2_B} \,, \nonumber  \\
\eta_4 &=& \frac{4\xi_B\lambda_{mfp}P}{9+\xi^2_B} \,.
\end{eqnarray}  
The analytical results are depicted by different curves in Fig. \ref{fig:eta}.
We see excellent agreements with our numerical results. From the formulas Eq. (\ref{eta0_moment}) it is clear
that $\eta_1/\eta$ goes to a constant of $4/3$ for large $\xi_B$ (large $eB$), while all the other shear viscosity
coefficients go to zero.

Furthermore, by equating Eqs. (\ref{Ieta0}) and (\ref{Ieta2}) to $\eta_0$ and $\eta_2$ in Eq. (\ref{eta0_moment})
we find that
\begin{eqnarray}
\label{c1}
\dfrac{V}{T}\int_0^\infty dt  \left\langle T^{12}({\bf r}, t)T^{12}(0,0)\right\rangle &=&  \frac{12\lambda_{mfp}P}{9+4\xi^2_B}\,, \\ 
\label{c2}
\dfrac{V}{T}\int_0^\infty dt  \left\langle T^{13}({\bf r}, t)T^{13}(0,0)\right\rangle &=&  \frac{12\lambda_{mfp}P}{9+\xi^2_B}\,.
\end{eqnarray}
Therefore, the correlation function $\left\langle T^{12}( t)T^{12}(0)\right\rangle$ at $\xi_B$ (or $B$) is same as
$\left\langle T^{13}( t)T^{13}(0)\right\rangle$ at $2\xi_B$ (or $2B$). This behavior can be observed in 
Fig.~\ref{fig:correlation0}. A similar scaling behavior between $\left\langle (T^{11}(t)-T^{22}(t))T^{12}(0)\right\rangle/2$
and $\left\langle (T^{13}(t))T^{23}(0)\right\rangle$ seen in Fig.~\ref{fig:correlation3} can also be explained by
equating Eqs. (\ref{Ieta3}) and (\ref{Ieta4}) to $\eta_3$ and $\eta_4$ in Eq. (\ref{eta0_moment}).

\subsection{Electric conductivity coefficients}
\label{sec:level2}
Firstly, we calculate the longitudinal electric conductivity within BAMPS by applying the Kubo formulas Eq.(\ref{Ikappa1})
and Eq.(\ref{Ikappa1_hk}), respectively. $\sigma_\parallel$ is induced purely by an electric field, while 
$\kappa_\parallel$ is related to the diffusion (or heat transfer). The correlation functions that determine $\kappa_\parallel$
and $\sigma_\parallel$ are $\left\langle G^3(t)G^3(0)\right\rangle$ and $\left\langle j^3(t)j^3(0)\right\rangle$. We note that
these two correlation functions (also $\kappa_\parallel$ and  $\sigma_\parallel$) are not influenced by the magnetic field,
since the Lorentz force does not affect the dynamics in the direction of the magnetic field.  In Fig.~\ref{fig:gzgz} we show
the time evolution of the two correlation functions and see that $\left\langle G^3(t)G^3(0)\right\rangle$ decreases to zero,
while $\left\langle j^3(t)j^3(0)\right\rangle$ approaches to a non-zero value. The latter indicates an infinite large electric
conductivity $\sigma_\parallel$, which is true for a one-component system of charged particles, since there is no energy
loss of changed particles in each collisions. We mention that the electric conductivity of a multi-component system such
like the quark-gluon plasma has been calculated in \cite{Greif:2014oia} within BAMPS without a magnetic field. 

The electric conductivity $\kappa_\parallel$, which is related to the diffusion (or heat transfer), is finite. 
In the top panel of Fig.~\ref{fig:kappa},  $\kappa_\parallel$ scaled by the temperature $T$ are shown by the solid
symbols. Charges are multiplied out in the results with $e^2=4\pi /137$. The electric conductivity calculated here
is related with the diffusion coefficient by the Wiedemann-Franz law. In Ref. \cite{Denicol:2012cn} the diffusion
coefficients for a one-component system without a magnetic field are calculated by using the 
$14$, $23$, $32$, and $41$-moment Grad's method. To make comparisons, the results obtained in 
\cite{Denicol:2012cn}  are multiplied by $e^2/T$ according to the Wiedemann-Franz law and shown 
in the top panel of Fig.~\ref{fig:kappa} by the dashed and dotted line corresponding to the $14$ and $41$-moment
approximation. We see that our numerical results are about $17\%$ smaller than those in the $14$-moment
approximation, while they agree nicely with those in the $41$-moment approximation. 

We now turn to calculate the transverse and Hall electric conductivity. The time evolution of the correlation functions
corresponding to these conductivity coefficients [see Eqs. (\ref{Ikappat}), (\ref{Ikappax}), (\ref{Ikappat_hk}), and 
(\ref{Ikappax_hk})] are shown in Fig.~\ref{fig:gxgx} and Fig.~\ref{fig:gxgy}. The transverse conductivity coefficient
$\kappa_\perp$ and  $1/\rho_\perp$ scaled by $T$ are depicted in the middle panel of Fig.~\ref{fig:kappa} by the
solid and open symbols, respectively. The values of $1/\rho_\perp$ are divided by a factor of $100$. 
At $B=0$, $1/\rho_\perp$ is infinite [see Eq. (\ref{Ikappat_hk})], while $\kappa_\perp$ is finite.
Both are equal to the longitudinal electric conductivity $\sigma_\parallel$ and $\kappa_\parallel$, respectively. 
From the middle panel of Fig.~\ref{fig:kappa} we also see that both $\kappa_\perp$ and  $1/\rho_\perp$
become smaller for stronger magnetic field strength and $1/\rho_\perp$ is roughly $100$ times larger than $\kappa_\perp$.

In the bottom panel of Fig.~\ref{fig:kappa} we show the Hall electric conductivity $\kappa_\times$ and
$1/\widetilde{\rho}_\perp$ scaled by $T$. The latter is divided by a factor of $5$ for comparisons. 
At $B=0$, $1/\widetilde\rho_\perp$ is infinite [see Eq. (\ref{Ikappax_hk})], while $\kappa_\times$ is zero.
$1/\widetilde\rho_\perp$ agrees with the classical result $en/B$ when comparing with
the dotted curve in the bottom panel of Fig.~\ref{fig:kappa}. (Remember that $1/\widetilde\rho_\perp$ has
been divided by a factor of $5$.)  With increasing $B$, $\kappa_\times$ increases first and then decreases
as $en/(5B)$. Thus, $1/\widetilde{\rho}_\perp$ is almost $5$ times larger than $\kappa_\times$. 
We realize that the electric conductivity coefficients induced by an electric field are always larger that
those related with the diffusion (or heat transfer).

The anisotropic diffusion coefficients of a one-component system in a magnetic field have been calculated
in Ref. \cite{Denicol:2018rbw} by using the 14-moment Grad's method. We multiply these results by $e^2/T$
to obtain the electric conductivity coefficients according to the Wiedemann-Franz law:
\begin{eqnarray}
\label{kappa_moment}
\kappa_\parallel &=& \frac{3e^2\lambda_{mfp}n}{16T} \,,\,\, \kappa_\perp = \frac{48e^2\lambda_{mfp}n}{(256+225\xi_B^2)T} \,, \nonumber  \\ 
\kappa_\times &=& \frac{45e^2 \xi_B \lambda_{mfp}n}{(256+225\xi_B^2)T} \,.
\end{eqnarray} 
$\kappa_\parallel$ is exactly the same as that obtained from \cite{Denicol:2012cn} and has been shown
in the top panel of Fig.~\ref{fig:kappa}. $\kappa_\perp$ and $\kappa_\times$ from Eq. (\ref{kappa_moment})
are depicted by the dashed curves (scaled by $T$) in the middle and bottom panel of Fig.~\ref{fig:kappa}. We see
agreements with the numerical results.

\section{Conclusions}
\label{sec:conclusion}
In this work, we have calculated the anisotropic transport coefficients of relativistic fluids in the presence of
a magnetic field according to the Kubo formulas given in Refs. \cite{Huang:2011dc,Hernandez:2017mch}.
The time correlations of the components of the energy-momentum tensor and electric current, which are
fluctuating in time at thermal equilibrium, are calculated numerically within the kinetic transport approach
BAMPS. For comparisons with results from the early studies we have considered a massless one-component
Boltzmann gas with isotropic binary collisions, although calculations within BAMPS can be performed for
multi-component systems with more complicated scattering processes such like pQCD (in)elastic scatterings
of gluons and quarks \cite{Uphoff:2014cba}.

We have found that the magnetic field dependence of the five shear viscosity coefficients that we achieved
agrees perfectly with the analytical results obtained by using the $14$-moment Grad's approximations
\cite{Denicol:2018rbw}. For strong magnetic field $\eta_1$ approaches $4/3$-fold of $\eta$ (the standard
shear viscosity without the magnetic field) and all the other shear viscosity coefficients decrease to zero.

We have also compared two kind of electric conductivity coefficients with each other.
One electric conductivity coefficients are associated with the diffusion (or heat transfer), another coefficients
are induced by an electric field and have no cross effect with the diffusion constant (or heat conductivity).
We found that the three electric conductivity coefficients associated with the diffusion are always smaller than
those induced by an electric field. In addition, the magnetic field dependence of the three electric conductivity
coefficients associated with the diffusion agrees well with the results from the $14$-moment Grad's
approximations \cite{Denicol:2018rbw}. A better agreement for the longitudinal electric conductivity is seen
when comparing the result from the $41$-moment approximation \cite{Denicol:2012cn}.

The agreements between the numerical and analytical results on the anisotropic transport coefficients
for a one-component system of Boltzmann particles with isotropic scatterings confirm the general use of
the derived Kubo formulas for multi-component particle systems with more complicated scattering processes.
Calculations of the anisotropic transport coefficients for a multi-component system in a strong magnetic
field such like the QGP produced in heavy-ion collisions are in progress. New results will be shown in
a future publication.

\begin{acknowledgments}
The authors would like to thank Xu-Guang Huang and Dirk H. Rischke for fruitful discussions. 
This work was financially supported by the National Natural Science
Foundation of China under Grants No. 11575092, No. 11890712, and No. 11775123, 
and the Major State Basic Research Development Program
in China under Grants No.  2015CB856903.  C.G. acknowledges support
by the Deutsche Forschungsgemeinschaft (DFG) through the grant
CRC-TR 211 ``Strong-interaction matter under extreme conditions''.
The BAMPS simulations were performed at Tsinghua National Laboratory
for Information Science and Technology and on TianHe-1(A) at National Supercomputer Center
in Tianjin.
\end{acknowledgments}


\bibliography{References}

\end{document}